\documentclass[twocolumn,showpacs,preprintnumbers,amsmath,amssymb]{revtex4}
\usepackage{indentfirst}
\usepackage{hhline}
\usepackage{wrapfig, float}
\usepackage{array,longtable,tabularx}
\usepackage{graphicx}% Include figure files
\usepackage{dcolumn}% Align table columns on decimal point
\usepackage{bm}% bold math
\usepackage{subfigure}
\usepackage[small]{caption2}
\usepackage{csquotes}
\usepackage{footnote,enumitem}
\usepackage[referable]{threeparttablex}
\usepackage{boxhandler}
\usepackage{color}
\usepackage{hyperref}
\usepackage{textcomp}% qu
\renewcommand{\captionlabeldelim}{.}
\hypersetup{linkcolor=blue, citecolor=blue, urlcolor=blue, colorlinks=true}%, pdfborderstyle={/S 1}}
\begin{document}

%\begin{center}

%\preprint{1507.08513}

\title{Collective infrared excitation in rare-earth Gd$_x$La$_{1-x}$B$_6$ hexaborides}

\author{E.~S.~Zhukova$^{1,2}$}\author{B.~P.~Gorshunov$^{1,2}$}\email{bpgorshunov@gmail.com} \author{G.~A.~Komandin$^{2}$}\author{L.~N.~Alyabyeva$^{1}$} \author{A.~V.~Muratov$^{3}$} \author{Yu.~A.~Aleshchenko$^{3}$}
\author{M.~A.~Anisimov$^{2}$}\author{N.~Yu.~Shitsevalova$^{4}$}\author{S.~E.~Polovets$^{4}$}
\author{V.~B.~Filipov$^{4}$}\author{N.~E.~Sluchanko$^{2,1}$}

\affiliation{$\phantom{x}^1$ -- Moscow Institute of Physics and Technology, 141700 Dolgoprudny, Moscow Region, Russia}

\affiliation{$\phantom{x}^2$ -- Prokhorov General Physics Institute of the Russian Academy of Sciences, 119991 Moscow, Russia}

 \affiliation{$\phantom{x}^3$ -- Lebedev Physical Institute, Russian Academy of Sciences, 119991 Moscow, Russia}

\affiliation{$\phantom{x}^4$  --  Frantsevich Institute for Problems of Materials Science, National Academy of Sciences of Ukraine, 03680 Kiev, Ukraine}

\date{\today}
\begin{abstract}
Using Fourier-transform infrared spectroscopy and optical ellipsometry, room temperature spectra
of complex conductivity of single crystals of hexaborides Gd$_x$La$_{1-x}$B$_6$,
$x$(Gd)$=\textmd{0}$, 0.01, 0.1, 0.78, 1 are determined in the frequency range
30--35000$~\textmd{cm}^{-1}$. In all compounds, in addition to the
Drude free-carrier spectral component, a broad excitation is discovered
with the unusually large dielectric contribution $\Delta$$\varepsilon$=5000 -- 15000
and non-Lorentzian lineshape. It is suggested that the origin of the excitation
is connected with the dynamic cooperative Jahn-Teller effect of B$_6$ clusters.
Analysis of the spectra together with the results of DC and Hall
resistivity measurements shows that only 30--50$\%$ of the conduction
band electrons are contributing to the free carrier conductivity
with the rest being involved in the formation of an overdamped excitation,
thus providing possible explanation of remarkably low work function
of thermoemission of Gd$_x$La$_{1-x}$B$_6$ and non-Fermi-liquid behavior in GdB$_6$ crystals.
\end{abstract}

\maketitle

%\textbf{1.}
\section*{I. INTRODUCTION}

%%% ----------------------------------------------------------------------
Rare earth (RE) borides $R$B$_n$ ($R$ is a metal ion, $n$$=$2, 4, 6, 12, etc.) compose a large family of compounds that exhibit a broad variety of properties depending on the $n$$=$$[\textmd{B}]/[\textmd{R}]$ ratio and on the type of metallic atom hosted by boron cage (in compounds with $n$$\geq$$6$; for $n$$<$$6$ the boron atoms form a \textit{planar} network) \cite{1}. Remarkable mechanical, chemical and thermal characteristics determined by boron framework of the systems make them promising for various technical applications and stimulate the ongoing studies of their fundamental physical properties. In this respect, especially attractive are borides with polyhedral-shaped boron cages ($n$$\geq$$6$) with metal atoms residing within. It is the complicated interplay between large-amplitude rattling vibrations of these caged atoms, lattice dynamics, electronic, magnetic and orbital subsystems that is expected to be at the origin of specific phenomena observed in RE borides and distinguishing them from other materials. Collective interactions of this type are usually characterized by relatively low energies, typically of the order of tens of meV or below. An effective experimental means to study the origin of such interactions is provided by optical spectroscopy that is able to deliver information on single-particle and collective excitations associated with charge, spin, orbital and phonon degrees of freedom, on microscopic parameters of charge carriers and on the mixed-type phenomena involving different subsystems. It is worth noting that while, indeed, Raman experiments are actively performed to study optical response of borides, infrared (IR) spectroscopy, which is complementary to the Raman spectroscopy, has not been used so often, especially in the most intriguing range of low energies (frequencies). This is due to the high electrical conductivity of borides with $n$$\leq$$12$ (except the narrow-gap semiconductors SmB$_6$ and YbB$_{12}$) that makes traditional for the IR Fourier-transform spectroscopy reflectivity measurements rather difficult due to closeness of reflection coefficient to $100\%$. In addition, the free charge carriers are very effective in screening dipole moments associated with possible IR-active excitations making them hardly observable experimentally. As a result, the IR experiments are rather scarce and relate mainly to the investigations of high-energy interband transitions. To fill the gap in the studies of low-energy electrodynamic properties of RE borides, we have performed in \cite{2} measurements of broad-band infrared spectra of LuB$_{12}$, the dodecaboride with lutetium ion that is non-magnetic excluding additional complications due to magnetoelastic effects and strong electron correlations. Special measures taken for careful measurements below 10000$~\textmd{cm}^{-1}$, where the reflection coefficient of dodecaborides gets strongly enhanced , allowed us to discover a broad non-Lorentzian lineshaped IR excitation with unusually large dielectric contribution ($\Delta$$\varepsilon$$\approx$8000). We have associated its origin with cooperative dynamics of Jahn-Teller active B$_{12}$ complexes that produce quasi-local vibrations (rattling modes) of caged lutetium ions.

\begin{figure}[t]
\begin{center}
\includegraphics[width = 8cm]{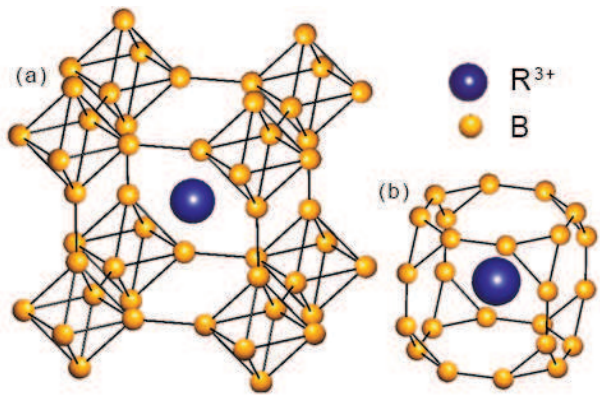}
   \parbox{8cm}{\caption{ (Color online). (a) Crystal structure of $R$B$_6$. (b) Fedorov B$_{24}$ polyhedra centered by $R^{3+}$ ion.
   }}\label{FigX1}
   \end{center}
\end{figure}

The aim of the present work is to apply similar approach for the search of low-energy excitations and thus for the study of the nature of the ground states of two representatives of RE metallic hexaborides LaB$_6$ and GdB$_6$ and their solid solutions Gd$_x$La$_{1-x}$B$_6$. $R$B$_6$ compounds crystallize in a \textit{bcc}-structure of CsCl-type with Pm3m-O$_1^h$ symmetry where loosely bound RE atom is located at Cs site and octahedral B$_6$ complex occupies the Cl position (Fig.\hyperref[FigX1]{1a}). As a result, there are two subsystems in the crystal structure, one of them is the rigid boron covalent network and another is loosely bound $R^{3+}$ ions embedded in cavities (B$_{24}$ cubooctahedra, see Fig.\hyperref[FigX1]{1b}) of the boron cage \cite{3,4}. For this reason vibrations of the heavy ions in $R$B$_6$ show the flat and low energy dispersion branches (Einstein oscillators with the temperatures $\Theta_E$(LaB$_6$)$\approx$140--150$~$K and $\Theta_E$(GdB$_6$)$\approx$91$~$K) \cite{5}--\cite{7} and the amplitude of these rattling modes in GdB$_6$ is among the most largest in the whole $R$B$_6$ family \cite{5}.

\begin{figure}[t]
\begin{center}
\includegraphics[width = 8cm]{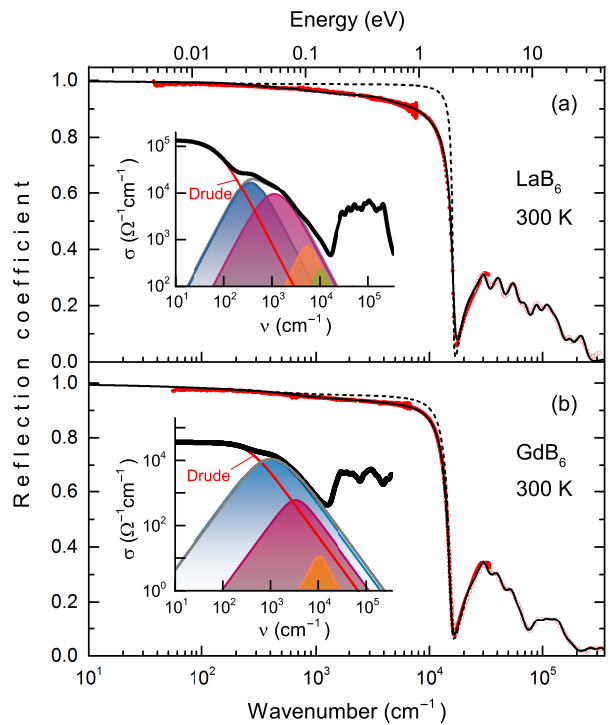}
  \parbox{8cm}{ \caption{(Color online). Room temperature reflection coefficient spectra of the LaB$_6$ (a) and GdB$_6$ (b) crystals. Dots show experimental data obtained using Fourier-transform spectrometer and ellipsometer, as described in the text. Open circles correspond to high-frequency reflectivity data from \cite{35}. Solid lines show the results of fitting the spectra using the Drude term, Eq.(\hyperref[Eq.1]{1}), for the free charge carrier response and Lorentzians, Eq.(\hyperref[Eq.2]{2}), responsible for absorption resonances. Dashed lines show best fits of the spectra that can be obtained using the Drude conductivity term alone with $\sigma_{DC}$$=$133330$~\Omega^{-1}\textmd{cm}^{-1}$,$\gamma^{Drude}$$=$330$~\textmd{cm}^{-1}$ for LaB$_6$ and $\sigma_{DC}$$=$35460$~\Omega^{-1}\textmd{cm}^{-1}$, $\gamma^{Drude}$$=$1050$~\textmd{cm}^{-1}$ for GdB$_6$. Insets show separately infrared contributions to the conductivity spectra (dots) from free carriers and Lorentzians.}}\label{FigX2}
   \end{center}
\end{figure}

Special interest to higher borides LaB$_6$ and GdB$_6$ is caused by two reasons. Firstly, both have an extremely low work function of thermoemission [$A$(LaB$_6$)$\approx$2.66$~$eV, $A$(GdB$_6$)$\approx$2.51$~$eV], \cite{8}, making one of them (LaB$_6$)  most commonly used material for thermionic cathodes in various electron-beam devices. It should be noted that the physical origin of the LaB$_6$ high thermoemission efficiency is not clear up to now, and this stimulates corresponding active studies (see e.g. \cite{9}). Secondly, these two compounds attract active interest of researches due to their exotic fundamental physical properties. Indeed, LaB$_6$ is a nonmagnetic reference compound in the $R$B$_6$ family which possess diverse and unusual electronic and magnetic properties, including homogeneous intermediate valence with topological Kondo insulating ground state in SmB$_6$ \cite{10, 11}, heavy fermionic behavior with unusual multipole magnetic ordering in CeB$_6$  \cite{12, 13}, complex magnetic ground state in PrB$_6$--HoB$_6$ antiferromagnets \cite{14}--\cite{17} and itinerant ferromagnetic behavior in EuB$_6$ \cite{18}. Wherein, depending on the rare earth or transition metal ion, the compounds can be narrow-gap semiconductors (SmB$_6$, YbB$_6$ \cite{19, 20}), semimetals (EuB$_6$ \cite{18, 21, 22}), antiferromagnetic metals (PrB$_6$--HoB$_6$ \cite{14, 17, 23, 24}) and superconductors (YB$_6$ \cite{25}). Among the antiferromagnetic hexaborides GdB$_6$ is an $S$-state system (for Gd$^{3+}$, L$=$0, S$=$7/2), but demonstrates a non-Fermi-liquid behavior of the resistivity $\rho$$\sim$$T$ \cite{24} and two successive antiferromagnetic transitions which are accompanied by simultaneous structural distortions \cite{16, 24, 26}.

\begin{figure}[t]
\begin{center}
\includegraphics[width = 8.5cm]{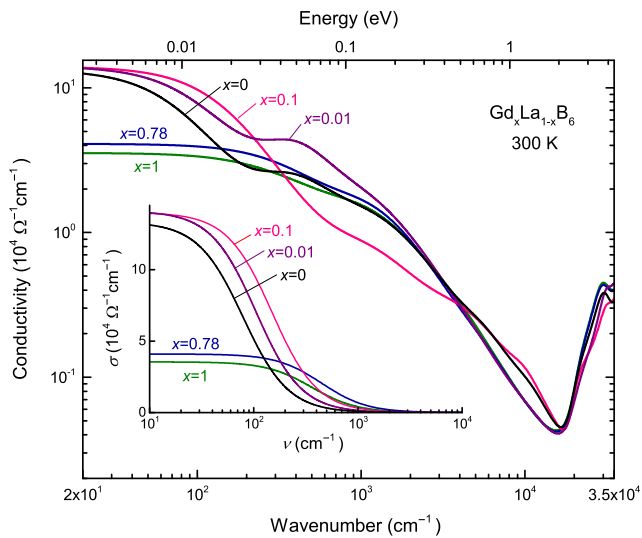}
   \parbox{8cm}{ \caption{(Color online). Room temperature spectra of real part of conductivity of Gd$_x$La$_{1-x}$B$_6$ crystals (solid lines). The spectra are obtained by least-square fitting of the reflection coefficient spectra using Drude conductivity term Eq.(\hyperref[Eq.1]{1}) and Lorentzian terms Eq.(\hyperref[Eq.2]{2}). Inset shows just the Drude conductivity spectra. }}\label{FigX3}
   \end{center}
\end{figure}

Such a variety of properties clearly indicates the existence of correlation effects whose optical fingerprints in the form of collective excitations are typically expected to exist at infrared and far-infrared frequencies. Anomalous peaks have been observed below 200$~\textmd{cm}^{-1}$ in the Raman spectra of $R$B$_6$ ($R$= Ca; Ce, Pr, Gd, Dy and Yb) \cite{27, 28} and attributed to a local rattling vibration of the $R$ ions in a shallow and unharmonic potential created by the boron cage. In \cite{29}--\cite{32} optical and infrared conductivity spectra of a series of hexaborides were measured in the energy range 1$~$meV to 40$~$eV. The observed excitations were analyzed and assigned to interband transitions. In \cite{30} it was noted that the obtained by Kramers-Kronig analysis optical conductivity spectra between $\approx$400$~\textmd{cm}^{-1}$ ($\approx$50$~$meV) and $\approx$8000 $~\textmd{cm}^{-1}$ ($\approx$1$~$eV) for a number of hexaborides (including LaB$_6$) could not be fitted with the simple Drude conductivity model. The effect was attributed to the electron-phonon and electron-electron scattering and modeled by introducing frequency-dependent scattering rate (and effective mass) of conduction electrons giving strong indication of collective interactions present in hexaborides.

To shed more light on the nature of such interactions, in the present study we have measured the room temperature broad-band infrared reflectivity spectra of Gd$_x$La$_{1-x}$B$_6$ single crystals with a series of gadolinium concentrations: $x$(Gd)$=$0 (LaB$_6$), 0.01, 0.1, 0.78, 1 (GdB$_6$). Exceptionally high quality of single-crystalline samples and extension of the measurement frequency interval down to as low as 30$~\textmd{cm}^{-1}$ (corresponding to the energy of $\approx$4$~$meV) made it possible to discover in these compounds overdamped infrared excitations characterized by the non-Lorentzian lineshapes. Comparative analysis of these collective excitations with that observed recently in a broad-band conductivity spectra of LuB$_{12}$ allows us to associate their origin with the dynamic cooperative Jahn-Teller effect of B$_6$ clusters. Detailed analysis of the free-carrier Drude conductivity spectra, together with the overdamped excitations and of their evolution with variation of $x$(Gd) allowed us to conclude that the Gd$_x$La$_{1-x}$B$_6$ hexaborides are \textit{non-equilibrium metals} where only 30--50$\%$ of electrons in the conduction band are contributing to the free carrier conductivity with the rest of the carriers involved in the formation of manybody complexes that in turn are responsible for strong changes in the 5$d$--2$p$ hybridization and hence for the modulation of the conduction band in the compounds.

\begin{figure}[t]
\begin{center}
\includegraphics[width = 8.2cm]{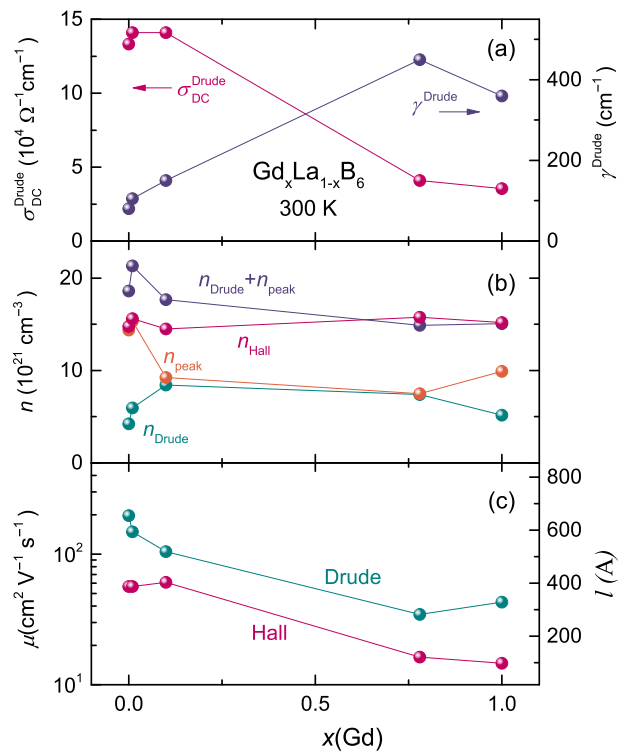}
\parbox{8cm}{\caption{ (Color online). (a) Dependences on the gadolinium concentration of charge carriers parameters that are responsible for the Drude conductivity in  Gd$_x$La$_{1-x}$B$_6$: DC conductivity $\sigma_{DC}^{Drude}$ and scattering rate $\gamma^{Drude}$. Panel (b) shows same dependences for free charge carriers concentration $n_{Drude}$, for concentration $n_{peak}$ of charge carriers responsible for the formation of collective absorption peaks, combined concentration $n_{Drude} + n_{peak}$ and the concentration $n_{Hall}$ of carriers obtained from Hall effect measurements. (c) Dependence on $x$(Gd) of charge carriers mobility and mean-free path obtained at room temperature from optical experiments and from Hall effect measurements. See also Table \hyperref[Tab.1]{I}.  }}\label{FigX4}
   \end{center}
\end{figure}

\section*{II. EXPERIMENTAL DETAILS  }\label{Sec.2}

\begin{figure}[t]
\begin{center}
\includegraphics[width = 8.05cm]{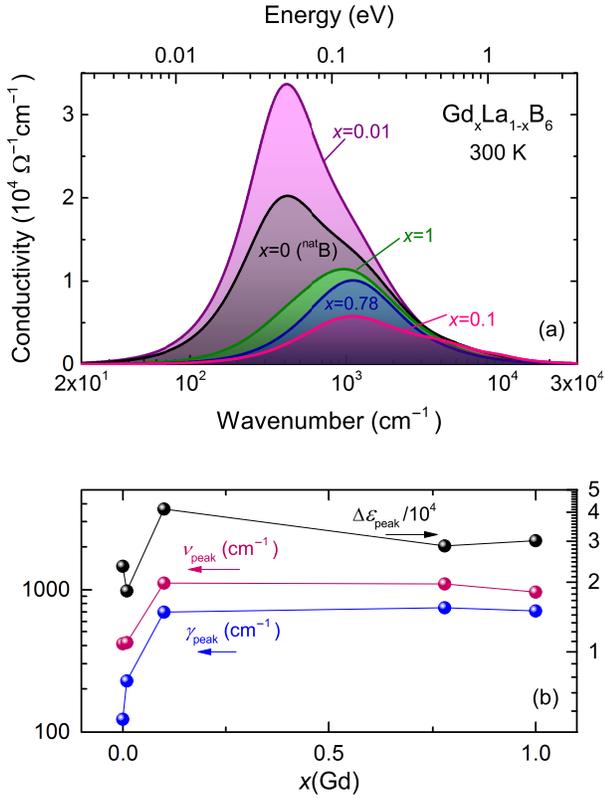}
   \parbox{8cm}{ \caption{(Color online). (a) Collective absorptions peaks observed in Gd$_x$La$_{1-x}$B$_6$ crystals that are modeled by a sum of Lorentzian terms, Eq.(\hyperref[Eq.2]{2}), as discussed in the text. (b) Dependence on the gadolinium concentration of the absorption peaks parameters: dielectric contribution $\Delta\varepsilon$, oscillator strength $f$ and damping constant $\gamma$. See also Table \hyperref[Tab.1]{I}. Temperature $T$$=$300$~$K.
 }}\label{FigX5}
   \end{center}
\end{figure}

High quality Gd$_x$La$_{1-x}$B$_6$ single crystals were grown by vertical crucible-free inductive zone melting in argon gas atmosphere on the setup described in detail in \cite{33}. The samples quality was characterized carefully by the X-ray diffraction, microprobe and optical spectral analysis, magnetization and DC transport measurements. For optical measurements $\approx$5$\cdot$5$~$$\textmd{mm}^2$ area samples were used with surfaces made plane within $\pm$1$~\mu\textmd{m}$ and polished with diamond powder. To avoid structural distortions on the surface, all samples were etched in the dilute nitric acid, according to a standard procedure \cite{34}. At frequencies $\nu$$=$30--8000$~\textmd{cm}^{-1}$, the reflection coefficient $R$($\nu$) spectra were measured using Vertex 80V Fourier-transform spectrometer. Gold films deposited on a glass substrate were used as reference mirrors. Using J.~A.~Woollam V-VASE ellipsometer, spectra of optical parameters [optical conductivity  $\sigma$($\nu$) and dielectric permittivity $\varepsilon'$($\nu$)] of the samples were directly determined in the interval 3700$~\textmd{cm}^{-1}$ -- 35000$~\textmd{cm}^{-1}$ with frequency resolution of 50$~\textmd{cm}^{-1}$. Measurements with the radiation spot diameter of 2$~$mm were provided with angles of incidence 65, 70, 75 degrees. From the ellipsometry data, reflection coefficients were calculated and merged with the measured infrared reflectivity spectra. The data from \cite{32} were used to extend the spectra up to $\approx$400000$~\textmd{cm}^{-1}$. The obtained broad-band reflection coefficient spectra were analyzed as described below. DC conductivities $\sigma_{\textmd{DC}}$ and Hall resistivity of the same samples were measured using a standard 5-probe method.

\section*{III.  RESULTS AND DISCUSSION}\label{Sec.3}

Fig.\hyperref[FigX2]{2} shows the reflection coefficient spectra of LaB$_6$ and GdB$_6$ that represent the two limiting compositions of the Gd$_x$La$_{1-x}$B$_6$ series. Features above 30000$~\textmd{cm}^{-1}$ (data from \cite{32}) are related to interband transitions and will not be discussed here. First of all, we note that although the overall $R$($\nu$) spectra look typically metal-like (there is a characteristic plasma edge at $\approx$17000$~\textmd{cm}^{-1}$ and the reflectivity approaches 100$\%$ at frequencies below $\approx$1000$~\textmd{cm}^{-1}$) they cannot be reproduced by the expression for the complex conductivity given by the Drude model:

\begin{equation}\label{Eq.1}
\sigma^*_{Drude}(\nu)=\frac{\sigma_{DC}^{Drude}}{1 - i\nu/\gamma^{Drude}} ,
\end{equation}

\noindent
where $\sigma_{DC}^{Drude}$ is the DC conductivity and  $\gamma^{Drude}$ is the charge-carrier scattering rate. This is demonstrated by the dashed lines in Fig.\hyperref[FigX2]{2} which show the best possible fit to experimental spectra using expression (\hyperref[Eq.1]{1}) alone. Similar mismatch between experiment and Drude fit was observed for all our Gd$_x$La$_{1-x}$B$_6$ samples with the deviations deceasing with the increase of $x$(Gd). To reveal the phenomena responsible for the detected effects we derived the spectra of optical conductivity by processing the reflection coefficient spectra with the Kramers-Kronig analysis. For high frequencies the $\nu^{-4}$ extrapolations were used. Towards low frequencies the spectra were extrapolated with the Hagen-Rubens expression $R$($\nu$)$=$$1-\sqrt{4\nu/\sigma_{DC}}$.  In addition, we performed spectral analysis of the $R$($\nu$) spectra by fitting them with expression (\hyperref[Eq.1]{1}) together with the minimal set (two to four dependent on the composition, see insets in Fig.\hyperref[FigX2]{2}) of Lorentzians needed to reproduce the measured $R$($\nu$) spectra

\begin{equation}\label{Eq.2}
\sigma^*(\nu)=\frac{0.5f\nu}{\nu\gamma+i(\nu_0^2 - \nu^2)} .
\end{equation}

\noindent
In Eq.(\hyperref[Eq.2]{2}) $\Delta\varepsilon$ is the dielectric contribution, $\nu_0$ is the resonance frequency, $f=\Delta\varepsilon\nu_0^2$ is the oscillator strength and  $\gamma$ is the damping constant. Both methods provided with basically same results that are presented in Fig.\hyperref[FigX3]{3} in the form of optical conductivity  $\sigma$($\nu$) spectra of all studied Gd$_x$La$_{1-x}$B$_6$ crystals. The spectra are shown for frequencies above 30$~\textmd{cm}^{-1}$ where they were essentially independent on various low-frequency extrapolations, and up to 35000$~\textmd{cm}^{-1}$ -- the highest frequency in our experiment. For clarity, we do not show in the figure the conductivity spectra directly measured with the ellipsometer.

%_______________ TABLE  I__________________
\LTcapwidth=16cm
\begin{longtable*}[b]{ccccccccc}%\firsthline
\caption{ Parameters of the Drude conductivity, Eq.(\hyperref[Eq.1]{1}), and the peak [composed by Lorentzians, Eq.(\hyperref[Eq.2]{2}), see text] terms used to describe the room temperature reflectivity spectra of Gd$_x$La$_{1-x}$B$_6$ crystals. Examples for spectra for LaB$_6$ and GdB$_6$ are presented in Fig.\hyperref[FigX2]{2}. (Drude term: $\sigma_{DC}^{Drude}$ -- DC conductivity,  $\gamma^{Drude}$ -- charge carriers scattering rate, $\nu_{pl}^{Drude}$ -- charge carriers plasma frequency, $n_{Drude}$ -- charge carriers concentration. Peak term: $f_{peak}$ -- oscillator strength, $\Delta\varepsilon_{peak}$ -- dielectric contribution, $\nu_{peak}$ -- peak maximum position, $\gamma_{peak}$ -- damping, $\gamma_{peak}/\nu_{peak}$ -- relative damping.) }\label{Tab.1} \\
\hhline{=========}\\
\quad\quad $x$(Gd) \quad\quad\quad & \quad $\sigma_{DC}^{Drude}$ \quad & \quad $  \gamma^{Drude}$ \quad & \quad   $(\nu_{pl}^{Drude})^2$ \quad &\quad $f_{peak}$ \quad & \quad $\Delta\varepsilon_{peak}$ \quad &\quad $\nu_{peak}$ \quad & \quad $\gamma_{peak}$ \quad & \quad $\gamma_{peak}/\nu_{peak}$ \quad\quad \\
\quad\quad  \quad\quad\quad & \quad ($\Omega^{-1}\textmd{cm}^{-1}$) \quad & \quad ($\textmd{cm}^{-1}$) \quad & \quad ($10^9 \textmd{cm}^{-2}$) \quad & \quad ($10^9 \textmd{cm}^{-2}$) \quad & \quad  \quad & \quad ($\textmd{cm}^{-1}$) \quad & \quad ($\textmd{cm}^{-1}$) \quad & \quad  \quad\quad \\
\\
\hhline{---------}\\
\quad\quad 0 \quad\quad\quad & \quad 133300 \quad & \quad 80 \quad & \quad 0.64 \quad & \quad 2.15 \quad & \quad 5130 \quad & \quad 416 \quad & \quad 1460 \quad & \quad 3.5 \quad\quad \\
\quad\quad 0.01 \quad\quad\quad & \quad 141000 \quad & \quad 105 \quad & \quad 0.89 \quad & \quad 2.29 \quad & \quad 7500 \quad & \quad 410 \quad & \quad 980 \quad & \quad 2.3 \quad\quad \\
\quad\quad 0.1 \quad\quad\quad & \quad 141000 \quad & \quad 150 \quad & \quad 1.27 \quad & \quad 1.38 \quad & \quad 14850 \quad & \quad 1110 \quad & \quad 3665 \quad & \quad 3.3 \quad\quad \\
\quad\quad 0.78 \quad\quad\quad & \quad 40980 \quad & \quad 450 \quad & \quad 1.1 \quad & \quad 1.1 \quad & \quad 15500 \quad & \quad 1100 \quad & \quad 2024 \quad & \quad 1.85 \quad\quad \\
\quad\quad 1 \quad\quad\quad & \quad 35460 \quad & \quad 360 \quad & \quad 0.77 \quad & \quad 1.48 \quad & \quad 15000 \quad & \quad 965 \quad & \quad 2207 \quad & \quad 2.3 \quad\quad \\
\\
\hhline{=========} \\
\end{longtable*}
%________________________ END  TABLE I______________________

In the main panel of Fig.\hyperref[FigX3]{3} one can clearly distinguish Drude-like conductivity increase towards low frequencies below 200--300$~\textmd{cm}^{-1}$ together with several absorption bands at $\nu$$\geq$200$~\textmd{cm}^{-1}$. The Drude free-carrier contribution to the conductivity spectra are shown separately in the inset. The dependence of the charge carriers scattering rate and of the DC conductivity on the gadolinium content is presented in Fig.\hyperref[FigX4]{4a} and in Table \hyperref[Tab.1]{I}]. Smooth and significant (by $\sim$5 times) growth of the scattering rate and corresponding decrease of the DC conductivity are observed when the concentration of Gd increases towards GdB$_6$. As to the absorption bands, they have rather unusual characteristics. Firstly, these bands are clearly seen in the reflectivity/conductivity spectra and are not completely screened by the charge carriers presented in the samples. Secondly, Lorentzian terms [Eq.(\hyperref[Eq.2]{2})] used to reproduce their spectral shape are characterized by relatively large values of oscillator strengths $f$$\sim$(1--2)$\cdot10^9$$~\textmd{cm}^{-2}$, dielectric contributions $\Delta\varepsilon$$\sim$5000--15000 and damping rates $\gamma/\nu_0$$\sim$1--3 (Table \hyperref[Tab.1]{I}). This means that the corresponding absorption mechanisms cannot be connected with regular phonons or interband transitions. We believe that the obtained overall conductivity spectra of all samples, in addition to the Drude components from free carriers, contain intensive excitations with pronouncedly non-Lorentzian lineshapes that was formally reproduced here by fitting the spectra with a sum of several terms described by Eq.(\hyperref[Eq.2]{2}). We show these excitations in Fig.\hyperref[FigX5]{5a} in the form of absorption peaks obtained by subtraction of the Drude components from the overall conductivity spectra. [We note that there are signs of weak bands around 10$^4~\textmd{cm}^{-1}$ that could be associated with interband transitions. Their oscillator strengths, however, are by one or two orders of magnitude smaller than those of the components that mainly determine the intensities of the peaks (see insets in Fig.\hyperref[FigX2]{2})]. We assume that the origin of the discovered overdamped excitations is connected with the Jahn-Teller (JT) instability of the B$_6$ complexes of natural boron (a mixture of $^{10}$B and $^{11}$B isotopes) which leads to emergence of cooperative dynamic JT effect in the studied compounds. We suggest that this kind of non-adiabatic mechanism launching both the cooperative overdamped modes and electronic instability could provide the most natural interpretation of the anomalies observed in the  $\sigma$($\nu$) spectra. It is seen from Fig.\hyperref[FigX5]{5} and Table \hyperref[Tab.1]{I} that with the growth of Gd content from $x$(Gd)$=$0 to $x$(Gd)$=$1, the collective absorption peak is blue-shifted by about 2.5 times and its dielectric contribution is nearly tripled. At the same time, the oscillator strength shows only slight tendency to decrease. Taking the value of the electronic effective mass $m^*$$=$0.6$m_0$ \cite{23} ($m_0$ is free electron mass), and using the relations for charge carriers plasma frequency $\nu_{pl}$$=$[$ne^2/(\pi m^*)$]$^{1/2}$  ($n$ is the concentration of charge electrons, $e$ -- their charges) and oscillator strength of the Lorentzians  $f$$=$$\Delta\varepsilon\nu_0^2$=$ne^2$($\pi m^*$)$^{-1}$ we can estimate the concentration $n_{Drude}$ of free carriers participating in the Drude conductivity and the concentration $n_{peak}$ of the charge carriers  involved in the formation of the overdamped excitations. The results are presented in Fig.\hyperref[FigX4]{4b}, together with the total concentration $n_{Drude} + n_{peak}$ that reveals only slight decrease when $x$(Gd) changes from 0 to 1. According to the estimates, for all studied compositions 50$\%$ or more of charges of the conduction band are involved in the formation of the collective excitation, with the highest fraction of more than $\approx$75$\%$ detected in LaB$_6$. Only the rest 25--50$\%$ of the carriers directly participate in the conduction process. It is worth noting that the total concentration, $n_{Drude} + n_{peak}$, found from optical measurements practically coincides with the concentration $n_H$$=$($R_He$)$^{-1}$ obtained from Hall effect measurements on the same samples ($R_H$ is the Hall coefficient) thus confirming the validity of the above estimates. The obtained values $n$$\sim$(1.5--2.1)$\cdot10^{22}~\textmd{cm}^{-3}$ coincide well with the electronic concentration obtained assuming one electron presented in the conduction band per unit cell, in accordance with the previously obtained results of transport measurements \cite{23, 36}. Furthermore, we can calculate mobility $\mu$$=$$e\tau/m^*$=e(2$\pi m^* \gamma^{Drude}$)$^{-1}$ and mean-free path $l$$=$$v_F\tau$ of the carriers responsible for the Drude transport [here $v_F$$\approx$6$\cdot$10$^7~\textmd{cm/s}$ is the Fermi velocity detected in \cite{37, 38}  and $\tau$$=$(2$\pi \gamma^{Drude}$)$^{-1}$ is the relaxation time]. Figure \hyperref[FigX4]{4c} demonstrates that the growth of gadolinium content leads to significant decrease of the mobility and mean-free path. Such behavior should be attributed to the carriers scattering on magnetic Gd ions and on rattling vibrations of Gd ions whose amplitude is significantly larger compared to that of La ions. It is interesting that the mobility values determined from the DC Hall effect measurements are much lower than those found from the AC optical study. This means that in the Hall experiments the mobility of all charge carriers is determined whose concentration is given by $n_{Drude} + n_{peak}$ including the concentration of the non-equilibrium conduction electrons with small mobility and mean free path, while the mobility values detected from optical measurements provide the mobility of Drude free electrons which are not involved in the formation of collective excitations.

The existence of cooperative JT dynamics of boron complexes that is suggested here to explain the origin of the discovered collective excitation in the $R$B$_6$ compounds induces immediately two effects. Firstly, the cooperative high-frequency JT boron vibrations provoke the rattling modes of heavy rare-earth ions which are quasi-local low-frequency vibrations whose equivalent Einstein temperature $\Theta_E$$=$90--150~K (dependent on the $R$-ion) is too small to be detected in optical spectra, but these are certainly detected in studies of the heat capacity and in the inelastic neutron scattering measurements \cite{5}--\cite{7}. Secondly, since the rare-earth ions are loosely bound in the rigid boron $R$B$_6$ cage, the Einstein oscillators are characterized by very large vibrations amplitude that leads to strong variation of the 5$d$--2$p$ hybridization of the $R$ and boron ions. As a result, modulation of the conduction band occurs producing "hot charge carriers" which are strongly scattered on the quasilocal mode.  In our opinion, the discovered large fraction of conduction electrons involved in the formation of collective excitation should be considered in terms of the non-equilibrium charge carriers thus providing possible explanation of remarkably low work function of thermoemission in the Gd$_x$La$_{1-x}$B$_6$ hexaborides. Moreover, these non-equilibrium and strongly scattered electrons in the hexaborides should be certainly taken into account when explaining the emergence of the non-Fermi-liquid regime of the charge transport in GdB$_6$ \cite{24, 26, 36}.

\section*{IV. CONCLUSIONS}\label{Sec.4}

Room temperature spectra of complex conductivity of single-crystalline Gd$_x$La$_{1-x}$B$_6$, $x$(Gd)$=$0, 0.01, 0.1, 0.78, 1 rare-earth hexaborides are determined in the frequency range 30--35000$~\textmd{cm}^{-1}$. It is demonstrated that the spectra contain two main contributions. The first is related to the response of free charge carriers and is described basing on the Drude conductivity model. In addition, an overdamped excitations with unusually large dielectric contributions $\Delta\varepsilon$$=$5000--15000 and oscillator strengths $f$$\sim$(1--2)$\cdot$10$^9 ~\textmd{cm}^{-2}$ are discovered whose origin is associated with the dynamic cooperative Jahn-Teller effect of B$_6$ clusters. The dependence on gadolinium content $x$(Gd) of parameters of both spectral components is determined and analyzed. It is shown that only 25--50$\%$ of the conduction band electrons are contributing to the free carrier conductivity with the rest being involved in the formation of the discovered collective excitation. The latter observation is supposed to be at the origin of remarkably low work function of thermoemission of Gd$_x$La$_{1-x}$B$_6$ crystals and may be responsible for emergence of the non-Fermi-liquid regime of the charge transport in GdB$_6$.

\section*{ACKNOWLEDGMENTS}

The research was supported by the Russian Science Foundation grant No.17--12--01426 and by the Ministry of Education and Science of the Russian Federation (Program 5 top 100). Authors acknowledge the Shared Facility Center at P.~N.~Lebedev Physical Institute of RAS for using their equipment.

%%%% ----------------------------------------------------------------------

\end{document}